\documentclass[iop]{emulateapj}
\usepackage{graphicx,amsmath,amssymb}
\usepackage{natbib}
\usepackage{apjfonts}
\bibpunct[, ]{(}{)}{;}{a}{}{,}

\newcommand{\sect}{\S}

\begin{document}
\title{Helium in natal \ion{H}{2} regions: \\ 
       the origin of the X-ray absorption in gamma-ray burst afterglows}

   \author{Darach~Watson\altaffilmark{1}}
   \author{Tayyaba~Zafar\altaffilmark{2}}
   \author{Anja~C.~Andersen\altaffilmark{1}}
   \author{Johan~P.~U.~Fynbo\altaffilmark{1}}
   \author{Javier~Gorosabel\altaffilmark{3}}
   \author{Jens~Hjorth\altaffilmark{1}}
   \author{P{\'a}ll~Jakobsson\altaffilmark{4}}
   \author{Thomas~Kr{\"u}hler\altaffilmark{1}}
   \author{Peter~Laursen\altaffilmark{1}}
   \author{Giorgos~Leloudas\altaffilmark{1,5}}
   \author{Daniele~Malesani\altaffilmark{1}}

   \altaffiltext{1}{Dark Cosmology Centre, Niels Bohr Institute, University of Copenhagen, Juliane Maries Vej 30, DK-2100 Copenhagen \O, Denmark;\\\texttt{darach@dark-cosmology.dk}}
   \altaffiltext{2}{Laboratoire d'Astrophysique de Marseille - LAM, Universit{\'e} Aix-Marseille \& CNRS, UMR 7326, 38 rue F. Joliot-Curie, 13388, Marseille Cedex 13, France}
   \altaffiltext{3}{Instituto de Astrof{\'\i}sica de Andaluc{\'\i}a (IAA-CSIC), Glorieta de la Astronom{\'\i}a s/n, 18008, Granada, Spain}
   \altaffiltext{4}{Centre for Astrophysics and Cosmology, Science Institute, University of Iceland, Dunhagi 5, 107, Reykjavik, Iceland}
   \altaffiltext{5}{The Oskar Klein Centre, Department of Physics, Stockholm University, 10691, Stockholm, Sweden}

   \begin{abstract}

Soft X-ray absorption in excess of Galactic is observed in the afterglows of
most gamma-ray bursts (GRBs), but the correct solution to its origin has not
been arrived at after more than a decade of work, preventing its use as a
powerful diagnostic tool.  We resolve this long-standing problem and find
that absorption by He in the GRB's host \ion{H}{2} region is responsible for
most of the absorption.  We show that the X-ray absorbing column density
($N_{\rm H_X}$) is correlated with both the neutral gas column density and
with the optical afterglow's dust extinction ($A_V$).  This correlation
explains the connection between dark bursts and bursts with high $N_{\rm
H_X}$ values.  From these correlations we exclude an origin of the X-ray
absorption which is not related to the host galaxy, i.e.\ the intergalactic
medium or intervening absorbers are not responsible.  We find that the
correlation with the dust column has a strong redshift evolution, whereas
the correlation with the neutral gas does not.  From this we conclude that
the column density of the X-ray absorption is correlated with the
\emph{total gas} column density in the host galaxy rather than the metal
column density, in spite of the fact that X-ray absorption is typically
dominated by metals.  The strong redshift evolution of $N_{\rm H_X}/A_V$ is
thus a reflection of the cosmic metallicity evolution of star-forming
galaxies and we find it to be consistent with measurements of the redshift
evolution of metallicities for GRB host galaxies.  We conclude that the
absorption of X-rays in GRB afterglows is caused by He in the \ion{H}{2}
region hosting the GRB.  While dust is destroyed and metals are stripped of
all of their electrons by the GRB to great distances, the abundance of He
saturates the He-ionising UV continuum much closer to the GRB, allowing it
to remain in the neutral or singly-ionised state.  Helium X-ray absorption
explains the correlation with total gas, the lack of strong evolution with
redshift as well as the absence of dust, metal or hydrogen absorption
features in the optical-UV spectra.

   \end{abstract}
   \keywords{ gamma-ray burst: general --- early universe --- dark ages,
              reionization, first stars --- galaxies: ISM
             }

   \maketitle

\section{Introduction}
\label{introduction}

Long-duration gamma-ray bursts (GRBs) accompany the deaths of some
stripped-envelope massive stars \citep{2003Natur.423..847H,2003ApJ...591L..17S}.  And while they have
been acknowledged as excellent probes of both the typical interstellar
medium of their host galaxies and foreground absorbers
\citep[e.g.][]{2004A&A...427..785J}, their use as probes of the progenitor star's
circumstellar material has met with limited success.  The main difficulty in
using GRBs to examine their immediate environs is that they are so luminous
across all wavelengths that they can be expected to ionise gas and destroy
dust to very large distances from the burst
\citep{2000ApJ...537..796W,2001ApJ...563..597F,2003ApJ...585..775P}, potentially eradicating
all traces of the circumstellar environment from their spectra, except
potentially in hot gas.  The best hope for observing the
environment near the burst could therefore be at X-ray wavelengths.

From the early soft X-ray observations of GRBs, large
absorptions were observed, initially believed to be consistent with the
light from the afterglow making its way out of a molecular cloud and
suffering absorption from the metals in the molecular cloud
\citep{2001ApJ...549L.209G,2002ApJ...565..174R,2006A&A...449...61C}. 
However, it has become clear since then that the picture is not that
simple. Indeed, the distribution of X-ray absorbing column densities is
substantially larger than one would expect in such a scenario.  And it has
been shown that the X-ray absorption is not readily connected to either the
neutral hydrogen column observed in the optical whether corrected for
metallicity or not, to low-ionisation gas, or to the column of dust
\citep{2007ApJ...660L.101W,2010MNRAS.402.2429C,2011A&A...525A.113S,2011A&A...532A.143Z}.

As we progress in our understanding, the nature of the X-ray downturn at
soft X-rays seems to become less and less clear.  We still do not know how
it arises, and indeed we are beginning to question whether it is even due to
photoelectric absorption \citep{2007ApJ...663..407B}.  This question has been a decade in the making
\citep{2001ApJ...549L.209G,2002A&A...395L..41W} and is one of the
outstanding observational issues related to GRBs.  Many suggestions have
been made as to its origin, among them highly-ionised gas in the
intergalactic medium (IGM), the host galaxy molecular cloud, or
circumstellar material related to the GRB progenitor.  

If we can resolve the origin of the X-ray absorption, we may 
find out about the immediate environment of the GRB, and
hence know about its progenitor and the matrix in which it was created; or,
if recent suggestions are correct \citep{2011ApJ...734...26B}, resolve the
missing baryons problem by detecting the warm-hot intergalactic medium.

In \sect~\ref{sect:observations} we present the problematic interpretation
of the existing data and discuss how the X-ray column density is determined
and presented.  In \sect~\ref{sect:evolution} we present our results on the
correspondence between the X-ray absorbing column density and the gas and
dust columns. The origin of the X-ray absorption is explored in
\sect~\ref{sect:whatisabsorber},
while its specific properties are discussed in
\sect~\ref{sect:basicproperties}. The implications of our findings are
examined in \sect~\ref{sect:implications} and our conclusions presented in
\sect~\ref{sect:conclusions}.

\section{Observational facts about X-ray absorbers in GRBs}
\label{sect:observations}

We now array the facts we know about the X-ray absorption in GRB afterglows.

In the Galaxy, there are strong correlations between the neutral hydrogen column density,
$N_{\rm H}$, the dust column density, $A_V$, and the X-ray absorbing column
density, $N_{\rm H_X}$
\citep{1978ApJ...225...40B,1978ApJ...224..132B,1985ApJ...294..599S,1994ApJ...427..274D,1998ApJ...500..525S,1975ApJ...198...95G,1995A&A...293..889P,2009ApJS..180..125R,2009MNRAS.400.2050G,2011A&A...533A..16W}.  The
proportionality constant is different in the Magellanic Clouds for
$N_{\rm H}/N_{\rm H_X}$, but corrected for the lower metallicities of these galaxies,
the ratio is quite similar
\citep{1982A&A...107..247K,1985A&A...149..330B,1985ApJ...299..219F,1985ApJS...59...77F,1986AJ.....92.1068F,1989A&A...215..219M,2003ApJ...594..279G,2008A&A...484..205D,2012ApJ...745..173W}.  The metals-to-dust ratio,
$N_{\rm H_X}$/$A_V$, is relatively constant not only in local group galaxies
\citep{2012ApJ...745..173W}, but also in galaxies at cosmological redshifts
\citep[e.g.][]{2006ApJ...637...53D,2009ApJ...692..677D}.  However in GRB
afterglows, the comparison with \ion{H}{1} and $A_V$ shows that $N_{\rm H_X}$ is
significantly larger than expected, whether metallicity corrections are
included or not \citep{2011A&A...532A.143Z,2012MNRAS.421.1697C}.  $N_{\rm H_X}$/$A_V$ is typically an order of
magnitude larger in GRBs than observed in other galaxies with a wide
variation \citep{2006ApJ...652.1011W,2007ApJ...660L.101W,2010MNRAS.401.2773S,2011A&A...532A.143Z}.
Finally, it was recently shown that the $N_{\rm H_X}$/$A_V$ ratio evolves
strongly with redshift \citep{2012ApJ...754...89W}.  This is strange and so
far remains unexplained.  We explore this phenomenon below.

The consensus view has so far been that the bulk of X-ray
absorption is not due to what is causing the dust
extinction and gas absorption observed in the optical and is probably
therefore due to very highly ionised gas
\citep{2007ApJ...660L.101W,2010MNRAS.402.2429C,2011A&A...525A.113S,2012MNRAS.421.1697C}.

\subsection{Practical considerations} 

The downturn observed in the soft X-ray spectra of GRB afterglows
\citep{2001ApJ...549L.209G,2002A&A...395L..41W} was initially assumed to be
the same as that observed in the Galaxy and in most extragalactic sources:
photoelectric absorption due to metals along the line of sight.  This
absorption is primarily due to the inner shell electrons of the most
abundant metals.  In the energy range observed by most modern X-ray
detectors (approximately 0.2--10\,keV) these metals are in particular C, N, O,
Ne, and S with contributions from the L-shell of Fe \citep[see for
example][]{2000ApJ...542..914W} and a contribution from He.  Oxygen
provides approximately 40\% of the opacity in the \emph{Swift}-XRT's
passband at $z=0$.  Because the absorption is caused by the inner shell
electrons, mildly ionised gas still absorbs soft X-rays, as do atoms in dust
or molecules, with relatively little alteration.  Thus, soft X-ray
absorption is a means to determine the total column density in metals in
front of an object.  Unfortunately, the low effective spectral resolution to
absorption features of all of our soft X-ray spectra is such that individual
features cannot be distinguished \citep[including the spectra obtained with
XMM-\emph{Newton} and \emph{Chandra}'s gratings, since the total numbers of
counts obtained with these instruments is so low,
e.g.][]{2003ApJ...587..128M,2003ApJ...597.1010B,2005ApJ...629..908B,2011MNRAS.418.1511C}.

\begin{figure}
  \includegraphics[angle=-90,bb=79 41 570 701,clip=,width=\columnwidth]{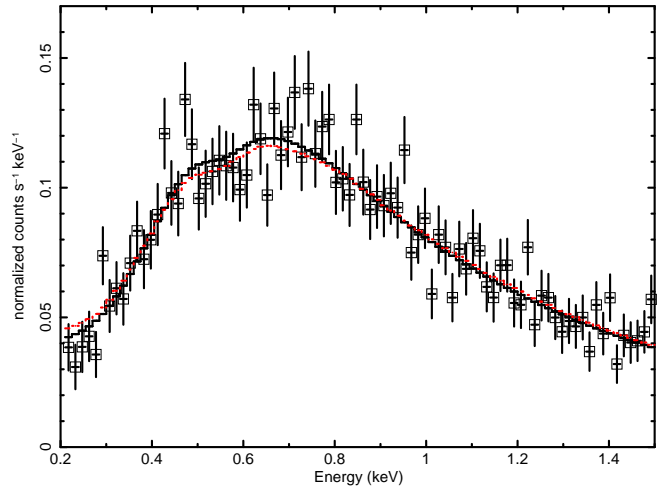}
  \caption{Soft X-ray spectrum of the afterglow of GRB\,121027A, convolved with the
           XMM-\emph{Newton} instrument response. The best-fit models for a
           power-law with a neutral, solar metallicity absorber (black, solid
           line), and a pure He absorber (red, dashed line) both at
           $z=1.773$ are shown. The models are practically
           indistinguishable.
   \label{fig:He_spectrum}}
\end{figure}

It's worth noting here that at low metallicities, the absorption is
increasingly dominated by He (and to a lesser extent, H, if it exists
in the neutral phase), and there is no clear way to distinguish absorption
by different elements.  As an illustration, in Fig.~\ref{fig:He_spectrum},
we show the spectrum of the X-ray afterglow of GRB\,121027A at $z=1.773$
\citep{2012GCN..13929...1T,2012GCN..13930...1K}, fitted with a traditional
solar abundance absorber, and with a pure He absorber.  The difference
between the fits for the models is negligible.  However, the total column
density in He is nearly a factor of seven larger for the pure He absorber. 
For singly-ionised He, this is another factor of $\sim25\%$ larger again,
since the cross-section for neutral He is somewhat larger in soft X-rays
than for He$^+$.  This means that for a \ion{H}{2} region, with a
considerable fraction of \ion{He}{2} but no metals, the gas column density
would be about 8 times the column required for a neutral, solar metallicity
gas at this redshift.

The lack of spectral resolution good enough to discriminate individual
features gives rise to complications and confusion in the presentation of
results.  The first is what column density to report.  Since we typically
not only do not know the state of the matter or indeed the precise elements
we are observing, a column density of any given species cannot be reported. 
For this reason an equivalent column density in hydrogen is usually reported
assuming, typically, a `solar' abundance of the elements---usually the
default abundance set in the software Xspec, from
\citet{1989GeCoA..53..197A}.  In spite of newer estimates of the solar
abundances that are about 40\% lower in metals
\citep[e.g.][]{2009ARA&A..47..481A}, it seems that the
\citet{1989GeCoA..53..197A} metallicity is a better approximation than
\citet{2009ARA&A..47..481A} for a typical Galactic sightline
\citep{2011A&A...533A..16W}.  For this reason and for ease of comparison
with previous estimates, it makes sense to continue to use
\citet{1989GeCoA..53..197A} metallicities unless we know what the absorber
is.  The second complication is that the absorbing column density is
occasionally reported as if the absorber was at $z=0$, typically where the
redshift is unknown.  The absence of spectral features means that the
redshift must be determined from observations at another wavelength,
typically optical or UV.  Since the absorption is shifted out of the
bandpass with redshift, the observed column density drops with redshift
approximately as $(1+z)^{-2.6}$.  Naturally the column density reported at
$z=0$ is thus substantially smaller than the actual value determined at the
correct redshift.  Third, the value found for the column density is almost
always determined assuming a neutral column density.  Ionised gas has a
smaller absorbing cross-section, and indeed, a different absorbing pattern.

\section{High X-ray absorption at high redshift: Why does $N_{\rm H_X}/A_V$ evolve with redshift?}
\label{sect:evolution}

It was noted by \citet{2010MNRAS.402.2429C} that there was a correlation
between the X-ray absorbing column densities in GRB afterglows and their
redshift, i.e.\ the highest redshift bursts were the most absorbed. 
\citet{2011ApJ...734...26B} demonstrated convincingly that this was not an
artifact of the fitting and assumptions, and suggested that this relation
was due to increasing foreground absorption by the highly-ionised metals
component of the intergalactic medium.  However, \citet{2012ApJ...754...89W}
found the ratio of the X-ray absorption to dust extinction $N_{\rm H_X}/A_V$
in GRB afterglows evolved with redshift, and that it was this evolution that
answered the mystery of the dearth of high X-ray absorbed GRBs at low
redshift.  This answer was in many ways more mysterious than the question. 
Why would the X-ray--determined metal column density increase relative to
the dust column density with redshift?  Here we examine the relationship
between $N_{\rm H_X}$ and $A_V$ as a function of redshift to understand the
origin of this peculiar evolution using (a slightly enlarged version of) the
data in \citet{2012ApJ...754...89W}.  These data are presented in detail in
Watson et~al.\ (in prep.).

\begin{figure*}
 \begin{center}
  \includegraphics[angle=-90,clip=,width=0.8\textwidth]{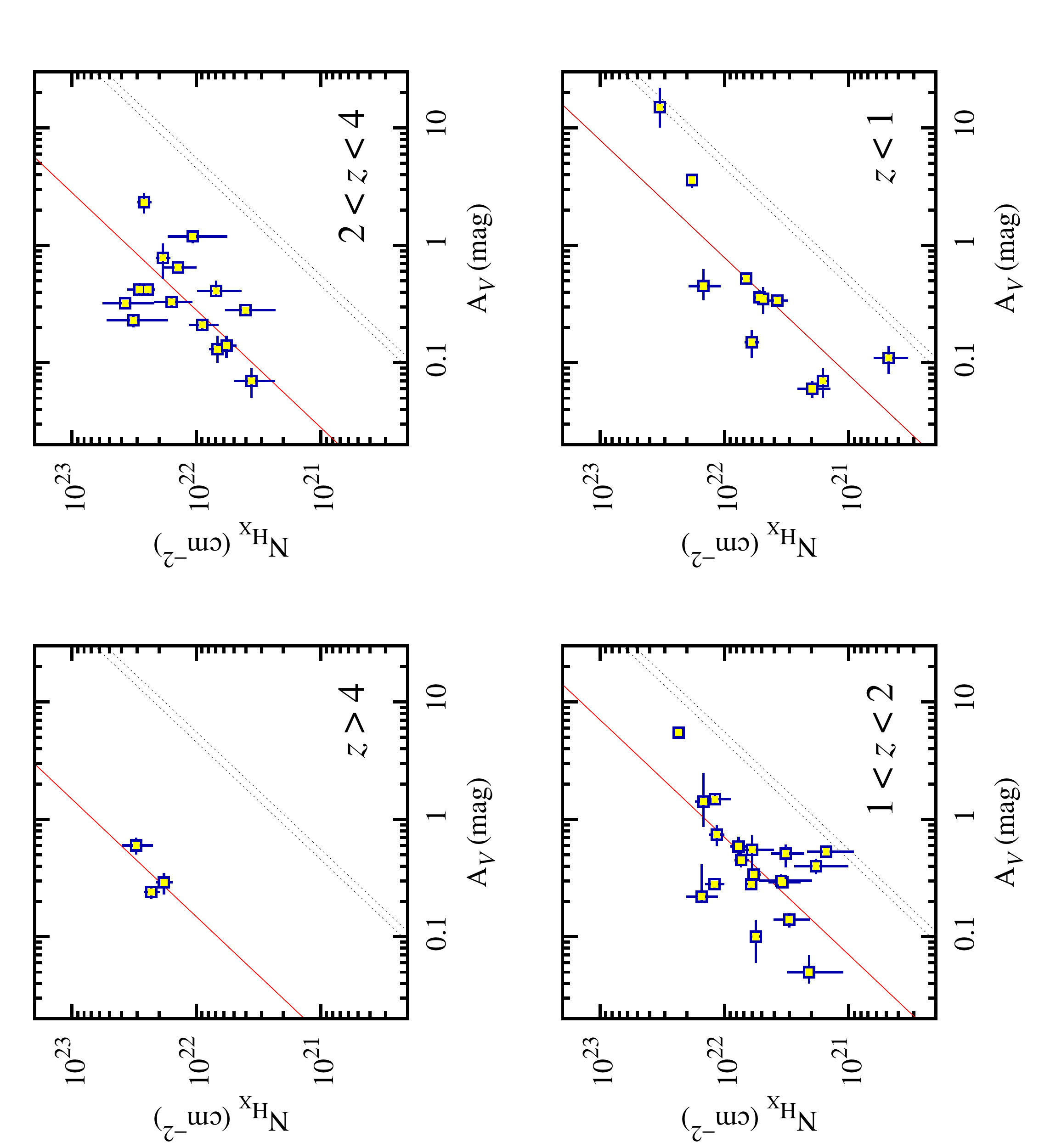}
  \caption{X-ray absorption versus extinction in GRB afterglows. The four
          panels present GRBs in different redshift intervals: \emph{top
          left} $z>4$, \emph{top right} $2<z<4$, \emph{bottom left} $1<z<2$,
          \emph{bottom right} $z<1$.  The solid line is the best-fit,
          fixed-slope line to the data.  The dashed lines mark the
          approximate limits of the metals-to-dust ratios reported for the
          local group of galaxies.  The metal and dust columns are clearly
          correlated in each redshift interval ($P>95\%$ in the three $z<4$
          datasets).  The best-fit metals-to-dust ratio rises at high
          redshift: from $\sim1\times10^{22}$\,cm$^{-2}$\,mag$^{-1}\,A_V$ at
          the lowest redshift to
          $\sim6\times10^{22}$\,cm$^{-2}$\,mag$^{-1}\,A_V$ at the highest. 
          The datapoints in faded colours are $N_{\rm
          H_X}-(2.0\times10^{21}$\,cm$^{-2}\times A_V)$ versus $A_V$.  The
          remaining X-ray absorption when the metal absorption corresponding
          to the near-universal metals-to-dust ratio (Zafar et~al., in
          prep.) is subtracted ($2.0\times10^{21}$\,cm$^{-2}\times A_V$) is
          still clearly correlated with $A_V$, showing that the $N_{\rm
          H_X}$--$A_V$ correlation we observe is not due to the neutral
          medium metals associated with the dust forcing a correlation at
          high $A_V$ values.}
  \label{fig:nx_av_z_panels}
 \end{center}
\end{figure*}

A close examination of Fig.~3 in \citet{2012ApJ...754...89W} hints at a
possible correlation between the metal column densities and the extinction
that evolved with redshift.  While there is no immediate correlation
apparent in the full dataset, splitting the data by redshift clarifies the
situation.  In Fig.~\ref{fig:nx_av_z_panels}, the distribution of X-ray
absorbing column density with extinction is plotted for GRBs in four
redshift intervals.  The correlation then becomes more apparent, with
correlation probabilities of greater than 99.9\%, 97.5\%, and 96\% for GRBs
in the intervals $z<1$, $1<z<2$, and $2<z<4$ respectively based on Kendall's
$\tau$.  The combined probability that such correlations occur randomly is
less than $1\times10^{-5}$.  Similar probabilities are obtained using
Spearman's $\rho$.  While the correlation is highly significant, there is
clearly significant scatter in this correlation. In sum: the X-ray
absorption is correlated with the dust extinction and the relation evolves
with redshift.

\begin{figure}
 \includegraphics[bb=50 61 275 332,width=\columnwidth,clip=]{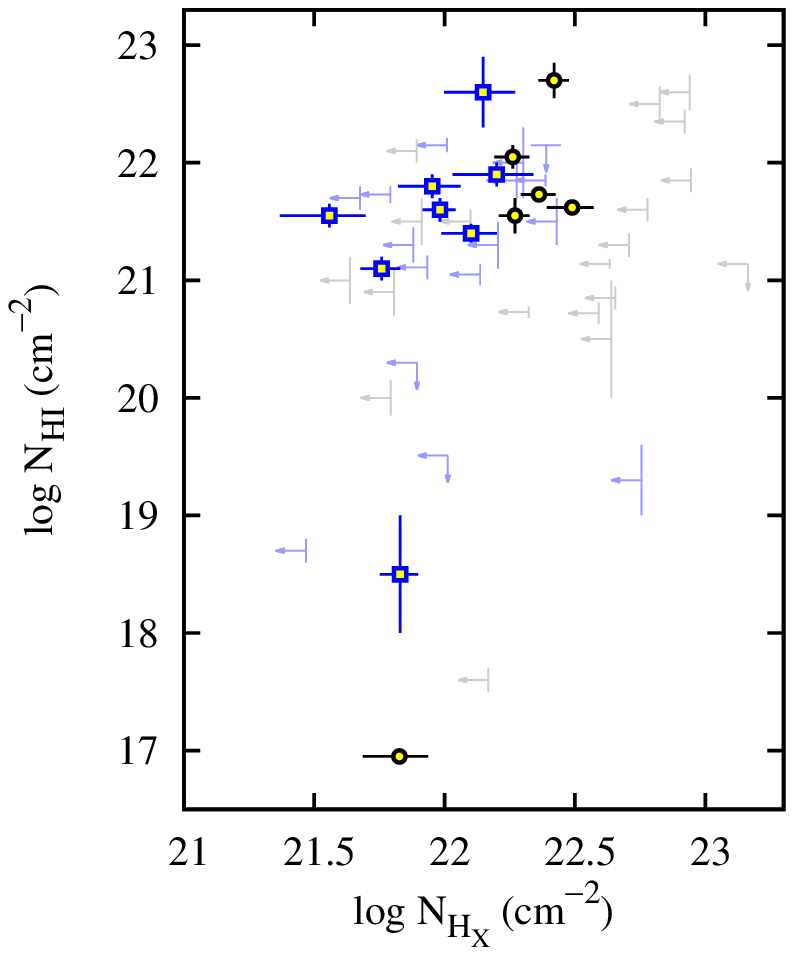}
 \caption{Gas (\ion{H}{1}) column density as a function of X-ray
          absorbing column density. Objects at high ($z\geq3$) and low
          ($z<3$) redshifts are marked with squares (blue) and circles
          (black) respectively.  At the higher values of $N_{\rm
          H\,\textsc{i}}$, i.e.\ for most objects, the gas and X-ray column
          densities are correlated, though with a considerable intrinsic
          scatter.  The correlation is similar to that observed between the
          dust extinction and the X-ray absorption
          (Fig.~\ref{fig:nx_av_z_panels}).  However, unlike the dust
          correlation, there is no obvious evolution with redshift.  Note
          that due to the low energy cut-off, the detectability threshold
          for the X-ray absorption is strongly redshift dependent.  This
          detectability effect is clear with the high redshift objects
          (black) having higher X-ray absorption and upper limits. 
          \label{fig:HIHX}}
\end{figure}

\subsection{Origin of the correlation between $N_{\rm H_X}$ and $A_V$}

At first glance, the fact that dark bursts are preferentially associated
with high $N_{\rm H_X}$ GRBs
\citep{2011A&A...534A.108K,2012MNRAS.421.1697C,2009ApJS..185..526F} implies
an $N_{\rm H_X}$--$A_V$ correlation.  The evolution with redshift and the
scatter however, was concealing it \citep[e.g.][]{2010MNRAS.402.2429C}.  The
first interpretation might be that this is simply host ISM in both dust and
metals, resulting in high $N_{\rm H_X}$ and $A_V$ values. However, we know that
the absorption in the optical-UV is dominated by low-ionisation gas and dust
at distances of hundreds of pc
\citep[see][]{2007ApJ...660L.101W,2007A&A...468...83V,2008A&A...491..189F,2009ApJ...694..332D,2010MNRAS.402.2429C,2011A&A...525A.113S,2013A&A...549A..22V}.
We also know that these low-ionisation metal column densities correlate
well with the observed $A_V$ \citep[Zafar et~al.\ in prep., see
also][]{2012A&A...548A..11D}. For the observed $N_{\rm H_X}$--$A_V$ correlation to hold, the cool, ISM
would have to dominate the column density. And indeed, it has been argued \citep{2011A&A...534A.108K} that even with both
components entirely uncorrelated, an approximate correlation between
\ion{H}{1} and $N_{\rm H_X}$ would be found at high column densities if the
\ion{H}{1} column is sufficiently dominant.

However this cannot be the origin of the correlations we observe with the
X-ray column.  First, the \ion{H}{1} column density, corrected for
metallicity, is almost never as large as the observed X-ray column density
\citep[see Fig.~\ref{fig:HIHX},
also][]{2011A&A...532A.143Z,2011A&A...534A.108K}, and the correlation
extends to low column densities.  We show in Fig.~\ref{fig:nx_av_z_panels}
the X-ray absorption values after subtraction of
$2.0\times10^{21}$\,cm$^{-2}\times A_V$ \citep{2011A&A...533A..16W}, which is the absorption expected for the
metals typically associated with such a column of dust (Zafar et~al.\ in
prep.).  The remaining X-ray absorption is still correlated with $A_V$, with
probabilities of 96\%, 94\% and 92\% for the three lower redshift sets.  The
combined probability of these datasets all having such strong correlations
by random chance is only $\sim3\times10^{-4}$.

Second, the correlation between the hydrogen column density and the X-ray
absorbing column density is tighter \emph{before} accounting for metallicity
\citep[see Figs.~3 and 4 in][]{2010MNRAS.402.2429C}.  This shows that the
correlation of the X-ray absorbing column density is primarily with the full
gas column density and not particularly with the neutral metals.  Third,
there is no obvious reason there should be an evolution in the $N_{\rm
H_X}/A_V$ ratio with redshift in this scenario.  We conclude, therefore that
the typical host galaxy ISM is not responsible for the correlation and that
\emph{there is a correlation between the dominant component of the X-ray
absorbing column density and the total gas column density.}

\subsection{High X-ray column density, high gas column density}

Such a correlation is odd at first, since the X-ray absorption is usually
principally due to metals.  However, what it is telling us, is not that the
gas column is responsible for absorbing the X-rays, but that where the X-ray
column density is high, so too, often, is the gas column density.  This
correlation with the gas column density then explains the strong redshift
evolution of the $N_{\rm H_X}$--$A_V$ relation, but the apparent lack of
redshift evolution of the $N_{\rm H_X}$--$N_{\rm H\,I}$ relation
(Fig.~\ref{fig:HIHX}).  It is simply the average metallicity evolving with
redshift.  At high redshifts we have high X-ray column density objects with
high gas column densities.  These objects are significantly metal poorer
than at low redshift, and hence (via the virtually constant dust-to-metals
ratio demonstrated in Zafar et~al.\ in prep.), have a lower $A_V$.  In other
words, the evolving $N_{\rm H_X}$--$A_V$ relation is a reflection of the
cosmic metallicity evolution.

We can now draw a few important conclusions from this. First, the X-ray
absorption component \emph{must} be associated with the host galaxy since it
correlates with the gas column density in the host.  Thus it cannot be due
to the intergalactic medium, for example.  Second, since the maximum values
of $N_{\rm H_X}$ are very similar at high and low redshift, and the mean
distribution certainly does not decrease to high redshift, as would be
expected if we were observing an X-ray absorbing column dominated by metals
in the ISM gas, it suggests that we are either observing metals ejected by
the progenitor star itself or that a component of the gas largely unaffected
by cosmic metallicity evolution is the absorber.

\subsection{The explanation of the `dark burst'--$N_{\rm H_X}$ connection}
The overall correlation between the X-ray column density and the dust
extinction can be explained by a model where GRB progenitors residing in the
hearts of galaxies are surrounded by higher density ISM.  Both \ion{H}{2}
regions and WR-nebulae are known to expand to smaller sizes in the hearts of
galaxies than those on the outskirts or outside the plane
\citep{2009A&A...507.1327H,2010MNRAS.409.1429S}.  These more compact sizes
result in higher observed column densities through the nebula or \ion{H}{2}
region.  GRBs occurring in high ISM density regions will therefore have
higher X-ray column densities.  Naturally, the afterglows of such bursts are
also much more likely to encounter high density ISM sightlines on their way
out of their hosts.  The approximate correlation observed between the X-ray
column density and the dust extinction Fig.~\ref{fig:nx_av_z_panels} is then
due to this effect.  The fact that `dark bursts' have on average higher
X-ray column densities
\citep{2011A&A...534A.108K,2012MNRAS.421.1697C,2009ApJS..185..526F,2012ApJ...758...46K} is
clearly an observational corollary.

We can confirm this explanation if we observe an overall and rough
correlation between the X-ray absorbing column density and the neutral
hydrogen column density.  This is the case.  In Fig.~\ref{fig:HIHX} we plot
the X-ray absorption against the neutral hydrogen column density determined
from the Ly$\alpha$ line.  There is an overall correlation between the two
(96\% confidence including all the data), with a few strongly discrepant
outlying objects with very low \ion{H}{1} column densities---which in this
scenario are objects that happen to find a clean line of sight out of the
host.  In an earlier, smaller sample, we found no evidence for a correlation
between the X-ray and H Ly$\alpha$ column densities, partly due to these few
very low H Ly$\alpha$ column density objects \citep{2007ApJ...660L.101W}.

The mysterious drop in the observed dust column for higher redshift sources
\citep[see][]{2012ApJ...754...89W} is readily explained in this context.  If
we assume that the average metallicity of galaxies is low at high redshifts
and the gas columns are approximately similar for similar X-ray absorptions
at all redshifts, and the dust-to-metals ratio is roughly constant, then the
apparent increase in the $N_{\rm H_X}/A_V$ is simply a result of the
decreasing metallicities at these high redshifts.  If this explanation is
correct, we should observe an evolution of increasing $N_{\rm H_X}/A_V$ with
redshift, but no such evolution in the $N_{\rm HI}/N_{\rm H_X}$ ratio. 
Again, this is what we observe, with no apparent evolution of the $N_{\rm
HI}/N_{\rm H_X}$ ratio (Fig.~\ref{fig:HIHX}).

A potentially fascinating tool resulting from this discovery is that the
observed evolution of the mean $A_V/N_{\rm H_X}$ for GRBs can therefore
yield an approximate indication of metal-enrichment of star-forming galaxies
with time.  At first glance, for example, we can argue that the metallicity
in star-forming galaxies decreases by approximately a factor of 3 between
$z<1$ and $2<z<4$ and by a factor of 5 between $z<1$ and $z>4$.
This is a flatter slope to the metallicity evolution than found with QSO-DLAs
\citep{2003ApJ...595L...9P}.  GRB host metallicities, on the other hand,
where they have been measured, also show an evolution with redshift, but the
evolution is significantly slower than for QSO-DLAs and is quite consistent
with what we observe here
\citep{2006A&A...451L..47F,2010AJ....139..694L,2012A&A...548A..11D,2012arXiv1206.2337T}. 
This is reassuring for the interpretation presented above.

\section{What is the X-ray absorber?}
\label{sect:whatisabsorber}

Given the correlation with gas in the host galaxy, it is clear that the
X-ray absorber must have some connection to the host ISM, but not the
typical free, low-ionisation ISM observed far from the burst. We are thus left
with two possibilities: the immediate cloud in which the burst progenitor
was born, or circumstellar material.

\subsection{The host molecular cloud}

It has been proposed that the X-ray absorption is related to gas and dust in
the molecular cloud in which the GRB progenitor is formed, but that the GRB
ionises the gas and destroys the dust to potentially large distances
\citep{2000ApJ...537..796W,2001ApJ...563..597F,2001ApJ...549L.209G,2003ApJ...585..775P},
leaving the X-ray absorption by highly-ionised metals as the only trace. 
However two pieces of evidence argue against this possibility.  First, the
low-ionisation optical absorber is observed to lie at very large distances
in almost every case, hardly ever at distances of a few pc
\citep{2007A&A...468...83V,2009ApJ...694..332D,2009A&A...503..437D,%
2011MNRAS.418..680D,2009A&A...506..661L,2008A&A...491..189F}. Molecular
hydrogen is very rarely detected
\citep{2009A&A...506..661L,2009ApJ...691L..27P,2008ApJ...682.1114W}, though
there is a strong dust-extinction bias against such detections. And there is
no one-to-one correlation between the X-ray absorption and the neutral
hydrogen column density \citep{2007ApJ...660L.101W}.  This means that the
hydrogen related to the X-ray absorber must be dissociated and ionised. 
However there are not enough UV continuum photons in most GRBs to dissociate
and then ionise the hydrogen out to distances much greater than $\sim 1$\,pc
from the burst for typical inferred column densities
\citep{2000ApJ...537..796W,2007ApJ...660L.101W}.  Second, there is little
evidence for the moderately ionised species one might expect to exist, lying
in the zone between the highly-ionised X-ray absorber and the almost neutral
optical absorber \citep{2011A&A...525A.113S}.  Ultimately, the GRB is not
UV-luminous enough to ionise the hydrogen to super-pc distances, while the
molecular clouds in which massive stars form should often be larger than
this size, and we do not observe any of the signatures of such molecular
clouds, of their ionisation by the GRB, or of the remnant neutral/molecular
material we would expect if the GRB were `burning' its way out of its
molecular cloud \citep[see][]{2000ApJ...537..796W}.  We therefore conclude
that host molecular clouds cannot be the origin of the X-ray absorption in
the general case.

\subsection{The host \ion{H}{2} region}

We must be careful to distinguish between a molecular cloud and what once
was a molecular cloud, which, after only a few million years will be
substantially ionised by the massive stars formed in it, transforming it
into a \ion{H}{2} region, likely before the burst occurs.  We are then
discussing \ion{H}{2} regions rather than molecular clouds, and the
arguments related to ionisation of hydrogen by the GRB do not apply.  The
total metal column densities are in the right range to explain the observed
X-ray absorption, in \ion{H}{2} regions with sizes up to several tens of pc
in size \citep[Fig.~4,][]{2009A&A...507.1327H}.  However, the principle
objection to this scenario is the observation that the $N_{\rm H_X}/A_V$
ratio evolves with redshift, which cannot readily be explained in this
scenario unless the absorption is not due to the metals in the gas.

The absence of an observed decrease in the X-ray column density to high
redshift is a strong indication that the X-ray absorber is not related to
the metals in the general ISM of the host galaxy.  If it was, the overall
evolution of the cosmic metallicity would imply a decrease of two orders of
magnitude in the absorption to $z=8$.  If the X-ray absorption was
influenced by the mean ISM metallicity, we would anticipate a strong
decrease in the mean X-ray absorption.  If anything, the mean X-ray
absorption appears to increase with redshift
\citep{2012MNRAS.421.1697C,2012ApJ...754...89W} and this cannot be a
selection bias, since the total number of objects at high redshift would
become unfeasibly large if there was a missing population of low-absorption
systems at high redshift.  In addition, there should be no strong bias
against discovering (not measuring) low-absorption systems at high redshift. 
We cannot see how an absorber dominated by the mean metallicity of the host
galaxy can reproduce the observed high absorption systems at high redshift,
and therefore are compelled to exclude \emph{metals} in the host \ion{H}{2}
region as the origin of the X-ray absorption.  However, it is worth
considering whether the He in the host \ion{H}{2} region could provide
enough X-ray absorption.  In that case we would need to demonstrate that
\ion{H}{2} regions have the correct distribution of gas and dust column
densities, and radii, and we would have to explain the lack of observations
of moderately-ionised species.

\subsection{Ejecta from the progenitor}

It is natural enough, considering the association of GRBs with type Ic SNe
\citep{1998Natur.395..670G,2003ApJ...591L..17S,2003Natur.423..847H}, to
suggest that GRBs have heavily mass-losing stars (specifically, Wolf-Rayet
(WR) stars) as progenitors
\citep[e.g.][]{1993ApJ...405..273W,1999ApJ...520L..29C}.  Recent
observations of WR stars in the Galaxy and Magellanic clouds show that the
WR nebula, filled with the highly-ionised, metal-rich ejecta of the star,
has a very high X-ray column density, typically about a few
$\times10^{22}$\,cm$^{-2}$
\citep{1994PASJ...46L..93K,2010AJ....139..825S,2011ApJ...727L..17Z,2011A&A...527A..66G,2012ApJ...747L..25O},
that is dust-free (or at least very dust-poor compared to the typical ISM).
The typical sizes of WR nebulae (a few pc) are consistent with the maximum sizes
estimated for the ionised GRB absorber \citep[though under a slightly
different set of assumptions]{2007ApJ...660L.101W,2011A&A...525A.113S}. 
The wind is already quite hot before the burst and so the bulk of the
absorbing material is already significantly ionised and may be far enough
away from the star that we would rarely if ever expect to see substantial
transient ionisation by the GRB itself.

We need not confine our considerations to mass-loss scenarios as observed
for known, well-observed Galactic WR stars.  It appears from recent work
that many evolved massive stars suffer significant mass loss in outbursts,
the archetype of which is $\eta$\,Carinae, where the mass loss from the
1840s event has been estimated to have been as much as 15\,M$_{\sun}$
\citep{2003AJ....125.1458S}.  These outbursts come, presumably, from stars
transitioning via precisely this mass-loss mechanism to the WR phase
\citep{2006ApJ...645L..45S}.  The nebulae around these high-mass stars are
often dusty and very massive, and appear to be more common than previously
believed \citep{2010AJ....139.2330W,2010MNRAS.405.1047G}.  Furthermore,
such extreme mass loss episodes have been discovered in scattered light,
dust reheating and absorption observed via supernovae, suggesting that
eruptive mass loss is common in the last millenium right up to explosion of
the star
\citep{2010ApJ...725.1768F,2007Natur.447..829P,2011MNRAS.418.1959S,2012ApJ...755..110C}.

The progenitor ejecta scenario explains many of the available observations,
including the high level of the X-ray absorbing column density compared to
$N_{\rm H\,\textsc{i}}$ and its relative lack of evolution with redshift.

However, in spite of the case made for progenitor ejecta, a major problem
remains with this interpretation: for distances of the absorber from the
progenitor larger than $\sim1$\,pc, the mass required becomes very large,
however the luminosity of the prompt GRB and early afterglow strips light
metals entirely of their electrons at anything except large distances.

This minimum distance is so large that for a given column density, the total
mass required is excessive for a progenitor wind.  Detailed analyses
\citep{2002ApJ...580..261P} as well as our own simple photon number
calculations, indicate that the GRB emission strips light metals to
distances of several tens of pc in some cases and a few pc even for lower
luminosity bursts.  But at distances of 1\,pc, as we note above, the mass
required to produce the mean observed GRB X-ray absorption is more than
$1\,M_\sun$ \emph{in metals} for spherically distributed ejecta.  This value
rises as the square of the distance, so that at a mean distance of even a
few pc, we would typically require approximately ten solar masses of light
metals.

These large masses essentially mean $r\lesssim1$\,pc for progenitor ejecta,
while in direct contradiction, the power of a typical burst completely
ionises metals that lie closer than a few pc, thus excluding an effective
absorber at distance less than a few pc \citep[e.g.][]{2002ApJ...580..261P}. 
Highly asymmetric mass loss, with the mass ejected along the direction of
the GRB, i.e.\ along the rotation axis of the star, could mitigate this
problem, but not enough and we still need the mass to lie at distances
typically smaller than $\sim1$\,pc.

\section{Properties of the absorber}
\label{sect:basicproperties}

We look at the properties listed above and examine specifically the
compatibility of these models with 1) the high level and distribution of
$N_{\rm H_X}$, 2) the evolving nature of the dust-to-$N_{\rm H_X}$ ratio,
3) the correlation between X-ray column density and the gas, 4) the large
typical distances to neutral gas, 5) the absence of moderately-ionised
species, and 6) any reports of features related to metals in the X-ray
absorber.

\subsection{\ion{H}{2} region or progenitor nebula?}

Both the progenitor ejecta and natal \ion{H}{2} region work well to
reproduce the observed features of the X-ray absorption. In particular, the
correlation between the X-ray absorbing column density and the gas column
density strongly suggests that the magnitude of the X-ray column is driven
largely by confinement by the host ISM, i.e.\ for a given mass, the column
density through it decreases as the square of the radius.  As observed for
WR nebulae and nearby \ion{H}{2} regions, the radius of the region is
related to galacto-centric distance \citep{2009A&A...507.1327H}, e.g.\ by
far the largest WR ejecta nebula is above the Galactic plane
\citep[W71,][]{2010MNRAS.409.1429S}.  For a given ejected mass, a physically
smaller bubble will result in a higher column density, giving rise to the
relation we observe in Figs.~\ref{fig:nx_av_z_panels} and \ref{fig:HIHX}.

However both hypotheses suffer from difficulties.  In the former scenario
there is a `mass-distance problem': a tension between the minimum distance
the metals must lie at in order not to be stripped of all their electrons by
the GRB and the maximum distance allowable to keep the mass required to
reproduce the observed column densities at a reasonable level.  We explore
possible solutions to the mass-distance problem in the appendix, but find no
solution consistent with the observed data and therefore reject it.  In the
latter scenario, the X-ray absorbing column density should drop with
redshift as the mean metallicity of star-forming regions drop.  A
He-dominated \ion{H}{2} region absorption would resolve this issue.  We
explore this scenario below.

\subsection{Properties of a helium-dominated \ion{H}{2} region absorber}
We examine here more quantitatively the prospect that the host \ion{H}{2}
region could have the properties required to reproduce the X-ray absorption. 

The reason He comes to dominate the X-ray spectra of GRB afterglows is
because the stripping radii related to a GRB are, somewhat
counterintuitively, smallest for H, then He and then the metals considerably
further out.  This is because the GRB emission has a far harder spectrum
than even hot stars; in photon flux terms, $F(E)\propto E^{-\Gamma}$,
where $\langle\Gamma\rangle \simeq 1$ \citep{2006ApJS..166..298K}.  The
number density of the atoms thus dominates the ionisation distances, and so
those elements with higher number densities have the smaller stripping
radii.  Therefore we would expect H and He to feature prominently in the
afterglow absorption, especially in low metallicity environments where GRBs
are found \citep{2008AJ....135.1136M,2012arXiv1211.7068G}.  However, in
\ion{H}{2} regions, the hydrogen is of course pre-ionised by the stars,
while the 54\,eV photons required to strip He are extremely rare, making He
the dominant absorber.  Indeed, we would expect a priori that He should
absorb GRB afterglows very strongly and it's worth asking the question: if
anything else were responsible for the excess X-ray absorption, then why
don't we see He absorption?

\begin{figure}
 \includegraphics[bb=3 4 412 348,width=\columnwidth,clip=]{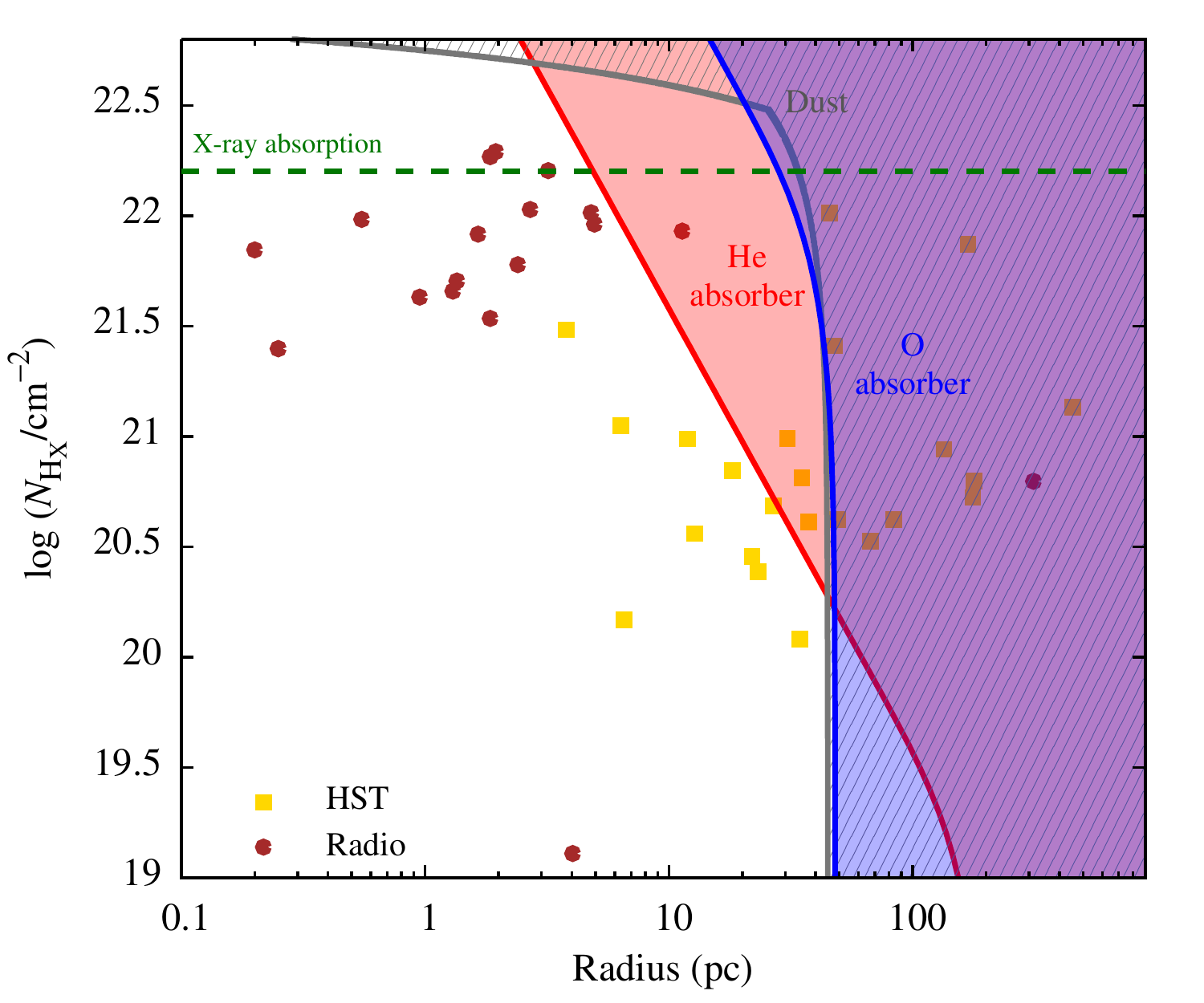}
 \caption{Equivalent hydrogen column density ($N_{\rm H_X}$) as a function of
          radius for a sample of nearby extragalactic star-forming regions
          from \citet{2009A&A...507.1327H} including many blue compact dwarf
          galaxies derived from radio observations (circles) and HST imaging
          with ground-based spectroscopy (squares).  The distribution of the
          radio-derived column densities, while containing possibly severe
          selection biases, is compatible with the mean observed $N_{\rm
          H_X}$ in GRBs \citep{2012ApJ...754...89W,2012MNRAS.421.1697C},
          indicating that the bulk of the X-ray absorption in GRBs could
          arise from material in its surrounding \ion{H}{2} region.  The
          column density for GRB\,050401 is indicated by a dashed line.  The
          white region indicates the distance to which GRB\,050401 has
          enough soft X-ray photons to fully strip helium and metals so that
          no X-ray absorption would be observed from the \ion{H}{2} region
          with a radius smaller than this distance in an afterglow.  Helium
          is stripped to smaller distances than oxygen (or other metals). 
          Dust will be destroyed by thermal sublimation to large radii.  The
          survival distances of He, O and dust from GRB\,050401 are
          indicated by red, blue and shaded regions respectively. 
          There is a large region where only He retains its
          electrons, making it the likely dominant X-ray absorber for most
          GRBs.  \label{fig:HII_regions}}
\end{figure}

We can calculate very roughly the effective stripping radius of various
elements making somewhat simplistic assumptions equating the total flux of
available photons and the total cross-sections of the hydrogen- and
helium-like ions (i.e.\ with only one or two electrons remaining) at these
radii.  We have done this for a representative, well-studied GRB,
GRB\,050401 at $z=2.9$ \citep{2006ApJ...652.1011W} and show these values in
Fig.~\ref{fig:HII_regions}.  It is clear that the space where the ionisation
state of He remains largely unaffected is very significant compared to the
stripping radius for O (the stripping radius for Fe is similar to O, and is
further out for C and other metals), leaving He as the only absorber in the
region between about 5\,pc and $\sim 30$\,pc.  This marks the effective size
of the \ion{H}{2} region surrounding GRB\,050401.  It must be larger than
5\,pc for there to be any He to absorb, and must be smaller than $\sim
30$\,pc, so that the metals and dust do not start to contribute
substantially to the optical/UV spectrum.  This limit on the size and our
measurment of the column density also allows us to determine that the
density of the \ion{H}{2} region must be in the range
$10^3-10^4$\,cm$^{-3}$.  These numbers are consistent with the densities
found for nearby \ion{H}{2} regions, though we should note that GRB\,050401
was a very luminous burst.  In Fig.~\ref{fig:HII_regions} we plot
the effective X-ray column density expected from \ion{He}{2} for the
densities and sizes of \ion{H}{2} regions in the sample presented by
\citet{2009A&A...507.1327H}.  The mean columns for the radio-determined
densities and sizes are surprisingly close the mean columns found for GRB
afterglows ($\log N_{\rm H_X}\sim21.7$).  The column densities determined
from the HST observations are consistently lower and may be affected by the
resolution of the HST imaging and ground-based spectroscopy
\citep{2009A&A...507.1327H}.

We have also plotted the thermal sublimation radius for dust according to
the prescription of \citet{2000ApJ...537..796W}.  However, we note that in
that paper they assumed only photons in the range 1--7.5\,eV would
contribute to the dust heating since molecular and atomic hydrogen would
absorb the UV photons above this energy. We use the parameters for the GRB\,050401
burst from the \emph{Swift}-BAT automated
analysis\footnote{\url{http://gcn.gsfc.nasa.gov/notices\_s/113120/BA/}}, as well as all the photons up to
40\,eV, since there is no H$_2$ or \ion{H}{1}.  This more than doubles the
effective thermal dust sublimation radius.  The dust sublimation radius is
rather close to the O and Fe stripping radii, and dust is therefore unlikely
to play a role in the absorption by the \ion{H}{2} region, again, leaving
only He as the expected signature.  The outcome of these considerations
aligns comfortably with the fact that the optical/UV extinction observed in
GRB afterglows is consistent with the column density of low-ionisation
metals observed far from the burst (Zafar et~al.\ in prep.) and with the
fact that only very small column densities of moderately ionised metals are
found in the spectra of GRB afterglows \citep{2011A&A...525A.113S}.

While H Ly$\alpha$ is commonly observed in the optical spectra of
$z\gtrsim1.8$ GRB afterglows, absorption lines associated with He are much
more difficult to detect.  The only lines we could reasonably expect to
observe in absorption are far into the UV, at or above the He Lyman series. 
For singly-ionised He, this effectively means that the absorption is never
observable, since for $z\lesssim3$, the continuum photons in the \ion{He}{2}
Ly$\alpha$ region of the spectrum are absorbed by neutral hydrogen in the
Galaxy, and for $z\gtrsim3$, the continuum is absorbed by the He Ly$\alpha$
forest.  The proportion of \ion{He}{1} to \ion{He}{2} depends of course on
the hardness of the radiation field ionising the region.  However, for
neutral He, a space-based, UV spectrograph might find the \ion{He}{1}
resonance line related to the X-ray absorpton for GRBs at $1\lesssim
z\lesssim2.5$, dependent on the proportion of \ion{He}{1} present.

\section{Implications of helium absorption}
\label{sect:implications}

We have concluded above that the data on X-ray absorption from GRBs allow us
to exclude every scenario proposed to explain it to date with the exception
of He in the natal \ion{H}{2} region.  Specifically, 
\begin{enumerate}
\item[a)] the absence of a
strong decrease in the mean X-ray absorption with redshift strongly implies
that the X-ray absorber is not primarily due to metals in the ISM of the
host galaxy;
\item[b)] the correlation we find between $N_{\rm H_X}$ and $A_V$
implies an association between the X-ray absorber and the host galaxy;
\item[c)]
the evolution of the $N_{\rm H_X}/A_V$ ratio coupled with the apparent lack
of evolution of this ratio for the optical metals to dust ratio
(Zafar et~al.\ in prep.) suggests that the X-ray absorber is not due to the metals
in the dust that cause the observed $A_V$;
\item[d)] that the evolution of the dust column is
related to the metallicity evolution, while the X-ray absorber is largely
independent of this;
\item[f)] this in turn suggests that the X-ray absorber is due
to He in the natal \ion{H}{2} region or progenitor ejecta metals;
\item[e)] the ionising power of the GRB precludes metals from the progenitor
ejecta;
\item[g)] the correlation between the X-ray absorbing column density and
the dust extinction and the atomic hydrogen column density indicates that
the host galaxy ISM has an influence on the X-ray column density.
\end{enumerate}

We are thus left with a scenario in which the X-ray absorption is caused by
He absorption in the natal \ion{H}{2} region that is undetected at
UV/optical wavelengths and that is confined by the host galaxy ISM.  Every
other scenario proposed so far is excluded.  

This conclusion leads to some predictions about the observable properties of
the absorber.  First, as mentioned above, there might be a detectable
signature of neutral He resonance absorption for GRBs in the redshift range
$1\lesssim z\lesssim2.5$, depending on the temperature of the \ion{H}{2}
region.  Second, since the absorption is dominated by He, the absorption
observed in X-rays should be smooth, i.e.\ absorption edges or lines due to
metals should be weak.  The claimed detection of S and Ne absorption edges
in the afterglow of GRB\,090618 seems problematic in this respect
\citep{2011MNRAS.418.1511C}.  \citet{2011MNRAS.418.1511C} found in their
analysis that S and Ne were significantly overabundant (though without Si or
O being similarly overabundant, which might have been expected on nucleosynthetic
grounds), but that the general metal abundance found was low.  However, it may          
be possible that the Galactic foreground was underestimated: an abundance
low in metals (particularly in O) was assumed \citep{2000ApJ...542..914W}
for the Galactic foreground absorption, substantially lower than typical for
Galactic sightlines \citep{2011A&A...533A..16W}.  At the same time, the
\ion{Ne}{1} edge at $z=0.54$, the redshift of the GRB, is very close in
energy to the Galactic \ion{O}{1} edge.  Clearly, accounting for the
Galactic \ion{O}{1} edge is crucial to an accurate determination of the
$z=0.54$ \ion{Ne}{1} edge.  Furthermore, observationally,
\citet{2011MNRAS.418.1511C} found the S and Ne abundances to be strongly
correlated, such that if the Ne abundance was low, the S abundance would
also likely be low.  Therefore we believe that the detection of absorption
edges in GRB\,090618 may need to be revisited.  No other non-transient
absorption feature has been reported in GRB afterglows, which is perhaps
noteworthy in its own right, (though we are unaware of any study explicitly
searching for these features in low-redshift GRB afterglows).  Currently
therefore, the data appear to be consistent with a smooth X-ray absorption.
 
Fits performed previously to the X-ray spectra of GRB afterglows with solar
metallicity material may therefore be somewhat unreliable not just in the
absolute value of the derived equivalent hydrogen column density, as is
generally acknowledged, but also in shape, especially at low redshift where
the O absorption edge is especially prominent.  The very low redshift
GRB\,060218, which is strongly absorbed, has an evolving thermal component
reported \citep{2006Natur.442.1008C,2007MNRAS.382L..77G}.  The fits to this
thermal component may be significantly affected by a He-dominated absorber
instead of a solar metallicity absorber.  

In addition to thermal components in low redshift bursts, the conclusions
drawn about the metallicity of highly absorbed bursts, especially at high
redshift, based on the X-ray absorption assuming the absorption to be
dominated by metals, should also be revisited in light of our results
\citep{2007ApJ...654L..17C,2011A&A...527A.104W}.

This first successful interpretation of the X-ray absorption allows it to be
used for diagnostic purposes. Based on the luminosity of the burst and the
continued existence of neutral or singly-ionised He, we can
set lower bounds to the radii of the \ion{H}{2} region in which the burst
resides, as well as upper limits based on the fact that the metals must be
stripped by the burst. With such limits, we can then convert the observed
column density to volume densities for these regions. Ultimately, it should
be possible to calculate approximate distributions for the densities and
sizes of individual \ion{H}{2} regions where massive stars form across a
vast range of redshifts, currently even as far as GRB\,050904 at $z=6.3$.

\section{Conclusions}
\label{sect:conclusions}

We have shown that the X-ray absorption in GRBs is correlated with the dust
extinction and with the neutral hydrogen column and that the ratio of
$N_{\rm H_X}/A_V$ changes with redshift in a way similar to the metallicity
redshift evolution of GRB host galaxies.  This suggests that the magnitude
of the X-ray absorbing column density in GRBs is correlated with the gas
density in the host galaxy ISM, explaining the relationship found between
high $N_{\rm H_X}$ bursts and so-called `dark bursts'.  Using these new
findings, we excluded all models relating the X-ray absorber to anything
outside the host galaxy including the warm-hot IGM.  On the grounds that the
$N_{\rm H_X}/A_V$ changes with redshift, we concluded that only helium in
the host \ion{H}{2} region or metals ejected by the progenitor star could be
the primary X-ray absorber.  However, the ionising power of the GRB sets a
minimum distance for metals to retain any electrons and hence be effective
X-ray absorbers; this places a minimum mass on the metals required which is
too large for ejecta from the progenitor, excluding this hypothesis.  We
thus concluded that helium in the host \ion{H}{2} region causes most of the
X-ray absorption observed in GRB afterglows.  This conclusion allowed us to
set limits on the sizes and densities of the typical \ion{H}{2} regions in
which X-ray absorbed GRBs explode, $\gtrsim$ a pc and $\lesssim$ a few tens
of pc, with densities about $10^3$--$10^4$\,cm$^{-3}$, consistent with
observations of sizes and electron densities of \ion{H}{2} regions in the 
Milky Way and in nearby galaxies.

\begin{acknowledgements} 

We would like to thank Paul Crowther, Nial Tanvir, Davide Lazzati, Enrico
Ramirez-Ruiz, James Rhoads, and Peter Jakobsen for invaluable discussions. 
The Dark Cosmology Centre is funded by the DNRF.  GL is supported by the
Swedish Research Council through grant No.\ 623-2011-7117.  

\end{acknowledgements}

\bibliography{mnemonic,grbs}

\appendix
\section{Possible solutions to the mass-distance problem for progenitor ejecta}
\label{app:progenitorejecta}

The mass-distance problem could possibly be resolved if the afterglow
doesn't see the same absorption as the prompt emission, i.e.\ the material
absorbing the afterglow is somehow not flash-stripped by the GRB.  One
apparently plausible scenario is that the afterglow has a substantially
larger opening angle than the prompt emission.  In this case, while the
prompt emission may strip metals to very large radii, it does so along a
very narrow beam, while the majority of the afterglow emission travels
through H- and He-like--ionised material much closer to the jet head. 
Another possible solution might be to clump the ejecta.  Finally, we could
conceal the metals in the solid phase.  In this latter case, optical
absorption lines would not be observed.  However, we would then need to
resolve the lack of observed extinction and the dust destroying power of the
GRB, both of which might be resolved by placing most of the metals in large
grains.  We explore these possibilities below.

\subsection{Implications of different prompt/afterglow opening angles}

A prediction of the different opening angles of the prompt and afterglow
emission is that the prompt phase should have much lower column densities on
average than the afterglow.  Whether this applies to later X-ray flares is
unclear, as they may have the same opening angle as the jet.  Some fraction
of cases where Swift-XRT follow-up of a very long burst, or a burst with a
precursor is rapid enough, should show absorption clearly rising as the
burst transitions from the prompt to afterglow phases.  So far, however,
this has not been observed.  Indeed, the opposite has been reported for
GRB\,050904 \citep{2007A&A...462..565G,2007ApJ...654L..17C}, and in our data we do not see clear evidence for a negative
difference between prompt and afterglow absorption, in direct conflict with
the requirements of this hypothesis.  It therefore seems unlikely that the
GRB prompt and afterglow emission see very different columns of gas.

Other solutions worth considering for the mass-distance problem are that the
X-ray absorber is clumped into extremely high column density knots, or
extremely high volume density knots.  In the former case, the knots have
high enough column density that even at distances of 0.1\,pc, the burst is
not powerful enough to strip all the electrons.  The covering fraction must
be extremely close to unity, however, and for the highest column density
situations, the absorber must be close to the Compton optical depth.  The
extremely large covering fractions seem unlikely, making this
scenario interesting but low probability.  In the latter case, very high
volume densities could potentially allow recombination on a short timescale
(e.g.\ $\sim1$\,s).  Such high volume densities, even at 0.1\,pc, imply an
incredible relative thickness of the shell ($\sim10^{-10}$), indicating that
this scenario is essentially impossible.

\subsection{Large grain hypothesis}

X-rays remove the inner electrons from metals. Once this happens, the
electrons do not recombine in an observable time, even for naked nuclei. 
This is the problem we are trying to overcome.  By placing the metals in the
solid phase, the affected ion can recombine rapidly, and the energy in the
liberated electron can be dissipated and radiated away at longer
wavelengths.  This process is more effective in larger grains.  The question
is whether any substantial fraction of the dust can survive the intensity of
the GRB at distances of a fraction of a pc, since it has been shown
theoretically that the extreme radiation field of a GRB is sufficient to
destroy dust to very large distances
\citep{2000ApJ...537..796W,2001ApJ...563..597F,2002ApJ...580..261P}.  We
hypothesise that the metals responsible for the X-ray absorption could be
primarily in large ($\sim1\,\mu$m) dust grains.  First, such grains are more
robust to UV sublimation.  The UV heating is approximately proportional to
the surface area of the grain.  Larger grains have a higher volume to
surface ratio and are thus more robust to sublimation.  Second, such large
grains could solve the problem that we do not observe substantial reddening
associated with the X-ray absorber \citep{2003ApJ...585..775P}.  However, on
closer inspection, it appears they cannot.  The absorption in X-rays has the
characteristic spectral shape of neutral medium absorption.  Specifically,
the spectrum is not flat at low energies, indicating the covering fraction
is close to unity.  This means that if metals in large grains are
responsible for the X-ray absorption, the grains must have a covering
fraction close to unity.  For most dust materials, the absorption efficiency
is close to unity.  Even for highly-transparent materials with relatively
low absorption in the optical (e.g.\ diamond), and large grains, the
scattering efficiency is close to unity, which would lead to very strong
extinction of the optical and UV light.  We thus conclude that we cannot
find a way to reconcile the progenitor ejecta hypothesis with the observed
properties of GRB afterglows.

\end{document}